\documentstyle[aps,epsf,twocolumn]{revtex}
\begin{document}
\twocolumn[\hsize\textwidth\columnwidth\hsize\csname @twocolumnfalse\endcsname

\title{Effect of Ferromagnetic Spin Correlations on Superconductivity in Ferromagnetic Metals}
\author{K.B. Blagoev$^*$, J.R. Engelbrecht, and K.S. Bedell}
\address{Department of Physics, Boston College, Chestnut Hill, MA 02167}
\date{\today}
\maketitle
\begin{abstract}
We study the renormalization of the quasiparticle properties in weak 
ferromagnetic metals, due to spin fluctuations,
away from the quantum critical point for small magnetic moment. 
We explain the origin of the $s$-wave superconducting instability
in the ferromagnetic phase and find that 
the vertex corrections are small and Migdal's theorem is satisfied 
away from the quantum critical point.
\end{abstract}
]
More than thirty years ago Doniach\cite{Doniach64} and Berk and 
Schrieffer\cite{Berk and Schrieffer66} showed that,
in the paramagnetic phase, 
the phonon-induced, $s$-wave superconductivity in
exchange-enhanced transition metals is suppressed by ferromagnetic
spin fluctuations, in the neighborhood of the Curie temperature. 
At the same time a theory of superconductivity coexisting
with long-range ferromagnetic order was developed by Larkin and Ovchinnikov
\cite{Larkin and Ovchinnikov64}, and by Fulde and Ferrell\cite{Fulde and
Ferrell64} for magnetic-impurity-induced ferromagnetism in metals. 
Without experimental evidence for the coexistence of 
superconductivity and ferromagnetism, this theory 
has been only of academic interest.
It is generally accepted that ferromagnetism suppresses superconductivity
and the apparent contradiction between the above two pictures has not
been clarified.

On the experimental side,
recent advances has allowed for the investigation of the
quantum critical region in 
correlation-induced, weak ferromagnetic metals 
\cite{Lonzarich} as well as in some heavy-fermion compounds\cite{Stewart et
al.97}. When hydrostatic pressure is applied on a transition metal compound
such as MnSi or ZrZn$_2$, the Curie temperature can be driven down to zero at a
critical pressure. In the neighborhood of this critical pressure the
paramagnetic-ferromagnetic phase transition is driven by quantum critical
fluctuations. So far, experiments have failed to find superconductivity in
the paramagnetic phase of these compounds and as we argue below, the physics
close to the phase transition is not well understood.

These experiments have motivated us to study the 
{\it ferromagnetic} regime relatively close to the critical point which
is described as a highly correlated but weakly ferromagnetic metal.
In this investigation we extend the Doniach, Berk, Schrieffer (DBS) theory 
into the {\it ferromagnetic phase} of the transition metal compounds.
We develop a microscopic theory of the ferromagnetic state,
based on the interactions mediated by spin-fluctuations between the fermions,
and explain the microscopic origin of the unexpected $s$-wave 
superconductivity recently
predicted by us\cite{BEB98} on the basis of phenomenological considerations.
In doing so, we find that, in contrast to the paramagnetic case, 
ferromagnetic fluctuations {\it enhance} pairing correlations -- 
resolving the longstanding dilemma referred to in the opening paragraph.

Our starting point is the Stoner state which is a Hartree-Fock solution 
of some Hamiltonian,
below the mean-field ferromagnetic instability\cite
{Doniach74}. This state is a product of two Slater
determinants with an
electron mass possibly renormalized by the band structure. At this level of
approximation the correlations do not renormalize the mass. 
Although the Stoner state has non-zero magnetization, 
it is known that the Hartree-Fock approximation overestimates the exchange.
However, we will assume that fluctuations about this mean-field saddle
point do not completely destroy the ferromagnetic order.
We cannot over-emphasize that this starting point is {\it not}
perturbatively connected to the paramagnetic Fermi-Liquid state.
The next step is to include the correlations which produce 
weakly interacting quasiparticles with renormalized mass $m^{*}$
near the Fermi surface.
The renormalization in the neighborhood of the two Fermi surfaces is
described by the single particle Green's function 
\begin{equation}
G_\sigma (\vec{p},\omega)=\frac z{\omega-v_F(|\vec{p}|-p_\sigma
)+i\delta_\sigma[\vec{p}]}+G_{inc}
\end{equation}
which is diagonal with the quantization axis parallel to the $z$-axis. Here 
$\vec{p}$ is the three-dimensional momentum of the particle, $p_\sigma$ 
($\sigma =\uparrow $,$\downarrow$) is the Fermi momentum of the spin-$\sigma$
electrons, 
$v_F$ is the Fermi velocity, 
and
$\delta_\sigma[\vec{p}]\equiv\delta\times\mbox{sign}(|\vec{p}|-p_\sigma ),$ 
with $\delta$ an infinitesimal real number.
The quasiparticle properties are
hidden in the quasiparticle residue $z<1$ and the effective mass. In
principle the Fermi velocity and the quasiparticle residue also depend on
the spin index, but in the neighborhood of the phase transition, 
$p_{\uparrow }-p_{\downarrow }\ll p_{\uparrow }$,$p_{\downarrow }$ 
and they are
equal. In the case of weak ferromagnetic metals the incoherent part of the
Green's function describing the physics away from the Fermi surface
is a smooth function of 
$\vec{p}$ and $\omega$ which renormalizes the properties at the Fermi 
surface and 
introduces no new physics. The Green's function describes a system of 
quasiparticles with spontaneous magnetization given by 
Dzyaloshinskii's theorem\cite{Dzyaloshinskii64} 
\begin{equation}
m_0=\frac 1{12\pi ^2}(p_{\uparrow }^3-p_{\downarrow }^3)=\frac{n_{\uparrow }-
n_{\downarrow }}{2}\text{,}
\end{equation}
where $m_0$ is the
uniform, static magnetization and $n_\sigma$ is the
density of spin-$\sigma$ particles. For weak
ferromagnets the magnetization $m_0$ is proportional to the exchange
splitting $\Delta =p_{\uparrow }-p_{\downarrow }$. Here we have assumed that
all particles are in an eigen-state of the $z$ component of the spin
operator and for definiteness we will assume that $p_{\uparrow
}>p_{\downarrow }$. The low-energy excitations of the system described by
this Green's function are quasiparticle excitations in the neighborhood of
the two Fermi surfaces as well as collective spin excitations. The
spontaneously broken $SU(2)$ symmetry guarantees the existence of a massless
Goldstone mode\cite{Wagner66} described by the propagator 
\begin{equation}
D_G(\vec{q},\omega )=-\frac{\Delta N(0)v_F}{2}\frac{\omega_s (\vec{q})}{
(\omega +i\delta )^2-\omega_s^2(\vec{q})}\text{,}
\end{equation}
$N(0)$ is the density of states at the Fermi surface.
In the case of a ferromagnetic metal the magnetization is a conserved
quantity and the spin-wave dispersion is $\omega_s(\vec{q})=D\left| \vec{q}
\right| ^2$ where $D=v_F\Delta /p_F^2$ is the spin stiffness.
The longitudinal response of the system is
described by the propagator\cite{Dzyaloshinskii and Kondratenko76} 
\begin{equation}
D_l(\vec{q},\omega )=
-{N(0)p_F^2\over2}
{1\over\xi^{-2}+\left|\vec{q}\right|^2-i\pi p_F^2\omega/2v_F\left|\vec{q}\right|},
\end{equation}
where $\xi \sim m_0^{-1}$ is the correlation length.
The interaction of the quasiparticles with these collective spin excitations
can be described by the interaction\cite{comment} 
\begin{equation}
H_{\mbox{sf}}=
g_0\sum_{\vec{k}\vec{q}\alpha\beta}c_{\vec{k}\alpha }^{\dagger }\vec{
\sigma}_{\alpha \beta }c_{\vec{k}+\vec{q}\beta }\vec{S}_{-\vec{q}}\text{,}
\end{equation}
where $g_0$ is the bare momentum-independent coupling constant, $c_{\vec{k}
\alpha }^{\dagger }$ and $c_{\vec{k}\alpha }$ are the anticommuting, 
quasiparticle creation and
annihilation operators respectively, $\vec{\sigma}_{\alpha \beta }$ are the
Pauli matrices and $\vec{S}_{-\vec{q}}=<\sum_{\vec{p}\gamma\delta}
c_{\vec{p}\gamma }^{\dagger }\vec{
\sigma}_{\gamma\delta}c_{\vec{p}-\vec{q}\delta}>$ is the three component spin
fluctuation field. The vector field $\vec{S}_{-\vec{q}}$ is the average
magnetization at a particular wave vector. In the ferromagnetic phase
this average is different from zero, while in the paramagnetic phase it
is strictly zero. Nevertheless it has been used to describe magnetically 
enhanced paramagnetic metals, although it can be mathematically 
justified only in the ferromagnetic phase. 

Recently we have shown\cite{BEB98}, ignoring the vertex corrections, that
the self energy leading to the exact Green's function, Eq.~(2), is local and
leads to logarithmic dependence of the quasiparticle residue on the
magnetization. When the magnetization approaches the quantum critical point
the quasiparticle residue vanishes and the Fermi liquid theory breaks down.
At finite temperatures in the neighborhood of the Curie temperature the spin
fluctuations lead to a non-Fermi liquid specific heat $C/T\sim \ln
T$ consistent with recent experiments on MnSi and ZrZn$_2$ as well as on some 
of the heavy-fermion compounds \cite{Lonzarich,Stewart et al.97}.

Weak ferromagnetic
metals are very interesting because the gapless Goldstone mode coexists
with the longitudinal excitations which are gaped. The longitudinal
spin-fluctuation propagator Eq.~(4) is similar to the model susceptibility
(peaked at the $\vec{Q}=(\pi ,\pi )$ nesting vector) used in the theory of
antiferromagnetic metals. However, in our case the expression for the
susceptibility is rigorous, following from the poles of the 4-point vertex
at small momentum transfer\cite{Dzyaloshinskii and Kondratenko76}.

The first vertex corrections to the 3-point vertex in the weak ferromagnetic
metal, Fig.~(1), are 
\begin{eqnarray}
\Lambda _{\uparrow \uparrow }^{(1)}(p,p+k)=\Lambda _{\uparrow \uparrow
l}^{(1)}+\Lambda _{\uparrow \uparrow G}^{(1)}\text{,}
\end{eqnarray}
\begin{eqnarray}
\Lambda _{\uparrow \downarrow }^{(1)}(p,p+k)=\Lambda _{\uparrow \downarrow
l}^{(1)}+\Lambda _{\uparrow \downarrow G}^{(1)}\text{,}
\end{eqnarray}
where 
\begin{eqnarray}
\Lambda _{\uparrow \uparrow l}^{(1)} &=&ig_0^2\int dq\;G_{\uparrow
}(q)D_l(q-p)G_{\uparrow }(q+k)\text{,}  \nonumber \\
\Lambda _{\uparrow \uparrow G}^{(1)} &=&ig_0^2\int dq\;G_{\uparrow
}(q)D_G(q-p)G_{\downarrow }(q+k)\text{,}  \nonumber \\
\Lambda _{\uparrow \downarrow l}^{(1)} &=&ig_0^2\int dq\;G_{\downarrow
}(q)D_l(q-p)G_{\uparrow }(q+k)\text{,}  \nonumber \\
\Lambda _{\uparrow \downarrow G}^{(1)} &=&ig_0^2\int dq\;G_{\downarrow
}(q)D_G(q-p)G_{\downarrow }(q+k)\text{,}  \nonumber \\
dq &\equiv &\frac{d^4q}{(2\pi )^4}\text{.}
\end{eqnarray}
Here we have assumed an expansion of the full vertex 
\begin{equation}
z\Lambda _{\alpha \beta }(p,p+k)=1+\Lambda _{\alpha \beta }^{(1)}+...\text{
,}
\end{equation}
and $p$, $q$, and $k$ are 4-vectors.

It is important to distinguish the order of the small momentum-transfer and
energy-transfer limits. In the limit which defines the Fermi liquid
parameters through the 4-point vertex, 
the Ward identity 
\begin{equation}
\lim_{\omega \rightarrow 0}\lim_{\vec{q}\rightarrow 0}\Lambda _{\alpha \beta
}(p,p+q)=\left( 1-\frac{\partial \Sigma }{\partial \omega }\right) \delta
_{\alpha \beta }=\frac 1z\delta _{\alpha \beta }\text{,}
\end{equation}
shows that the vertex is proportional to the inverse quasiparticle residue.
The effective pairing potential in principle can be constructed from the 
3-point vertex with the requirement that the triplet scattering amplitude
is zero. In second order perturbation theory, however, the momentum 
independence of the self-energy and the vanishing of the triplet scattering
amplitude are incompatible and so far we have not been able to construct
a pairing potential with the above properties. Nevertheless, one can
see that the singlet scattering amplitude is attractive leading to
a pairing instability in the singlet channel\cite{BEB98}. Physically, the Pauli
exclusion principle keeps quasiparticles with the same spin apart, leading
to a negative charge depletion between them. This charge distribution 
attracts another quasiparticle with the opposite spin leading to the
singlet pairing.

The Ward identity which we mentioned earlier shows that the effective 
pairing is enhanced for
small magnetizations since $z^{-1}\sim \ln m_0$ and this enhancement 
is due to the longitudinal collective mode. 

In the physical limit where energy is
conserved, the corresponding Ward identity is 
\begin{equation}
\lim_{\vec{q}\rightarrow 0}\lim_{\omega \rightarrow 0}\Lambda _{\alpha \beta
}(p,p+q)=\frac{v_F}z\frac{dp_\alpha }{d\mu }\sigma _{\alpha \beta }^z
\end{equation}
where, $\sigma _{\alpha \beta }^z$ is the Pauli matrix, $v_F^0$ is the Fermi
velocity of the noninteracting Fermi gas, and there is no summation
over repeated indexes.

In calculating the vertex corrections we first set the
frequency to zero and then take the limit for the momentum. Because,
we are working in the broken symmetry phase a distinction must be
made for vertex corrections involving particles on one of the two
Fermi surfaces and vertex corrections involving particles on
different Fermi surfaces. In the former case the limit
\begin{equation}
\Lambda_{\sigma\sigma}(|\vec{p}|\rightarrow p_\sigma,
|\vec{p}|\rightarrow p_\sigma)
\end{equation}
while in the latter the limit
\begin{equation}
\Lambda_{\sigma\sigma^\prime}(|\vec{p}|\rightarrow p_\sigma,
|\vec{p}|\rightarrow p_{\sigma^\prime}+\Delta)
\end{equation}
must be taken. In both cases we use the spectral representation
for the propagators $D_{l/G}$
\begin{equation}
D_{l/G}(\vec{q},\omega)=\frac{2}{\pi}\int_0^\infty 
\frac{z ImD_{l/G}(\vec{q},z)}{z^2-\omega^2-i\delta}
\end{equation}
Using that
\begin{eqnarray}
G_\sigma(\vec{p}+\vec{q},\epsilon+\omega)
G_{\sigma^\prime}(\vec{p}+\vec{q}+\vec{k},\epsilon+\omega)= \nonumber \\
\frac{z}{v_F}\frac{G_\sigma(\vec{p}+\vec{q},\epsilon+\omega)-
G_{\sigma^\prime}(\vec{p}+\vec{q}+\vec{k},\epsilon+\omega)}
{|\vec{p}+\vec{q}|-|\vec{p}+\vec{q}+\vec{k}|-(p_\sigma-p_{\sigma^\prime})}
\end{eqnarray}
it is not difficult to obtain the expansion
\begin{eqnarray}
z\Lambda_{\sigma\sigma;l}(p_\sigma,p_\sigma)= \nonumber \\
1+\frac{g_0^2N^2(0)z^2}{16p_\sigma}\ln\frac{\pi^2p_\sigma^4+4\Delta^4}
{\pi^2p_\sigma^4+4(\Delta^2+p_c^2)^2}+\ldots
\end{eqnarray}
and
\begin{equation}
z\Lambda_{\sigma\sigma;G}(p_\sigma,p_\sigma)=
1+\frac{g_0^2N^2(0)z^2}{4}ln(1+\frac{\Delta p_c}{p_F^2})+\ldots.
\end{equation}
We have used a momentum cutoff $p_c$ reflecting the different
physics at very small distances. Very similar logarithmic
behavior can be seen in the vertex expansion of 
$\Lambda_{\sigma\sigma^\prime;l/G}(p_\sigma,p_\sigma^\prime+\Delta)$.
This implies that the self-energy is weakly momentum dependent close to
the phase transition. Therefore a 
local ferromagnetic Fermi liquid theory\cite{Engelbrecht and Bedell95} 
can be used to describe weak ferromagnetic metals in a regime where the
magnetization is sufficiently small, but away from criticality (since the
fluctuations in the critical regime are beyond the scope of Fermi
liquid theory). This confirms the $s$-wave pairing 
instability\cite{BEB98} in the ferromagnetic phase.

The above adiabaticity is a consequence of the smallness of the exchange
splitting $\Delta $ compared to the Fermi momentum 
$p_F\equiv(p_{\uparrow }+p_{\downarrow })/2$ and the 
smallness of the maximum spin
wave velocity $\omega _G$ compared to the Fermi energy $\epsilon _F$ and
leads to the validity of Migdal's theorem\cite{Migdal58}
for weak ferromagnetic metals.

In the DBS theory of spin-fluctuation-enhanced paramagnetic metals it is
argued that the sharply peaked static spin susceptibility 
\begin{equation}
\chi (0,0)=\frac{\chi _p}{1-N(0)V_c}
\end{equation}
close to the Curie point suppresses the $s$-wave pairing, because
ferromagnetic spin fluctuations act as an effective repulsive force between
electrons with opposite spins. Here the $\chi _p$ is the Pauli
susceptibility and $V_c$ is a pseudopotential. To see why the
spin-fluctuations have opposite effect in the ferromagnetic phase it is
convenient to write the factor $1-N(0)V_c$ in the denominator of the spin
susceptibility in terms of the Landau Fermi liquid parameter $F_0^a$. Then
in both paramagnetic and ferromagnetic phases the static spin susceptibility
is positive and in the ferromagnetic phase is\cite{BEB98} 
\begin{equation}
\chi (0,0)=\frac{N(0)/2}{\left| 1+F_0^a\right| },
\end{equation}
while in the paramagnetic phase is
\begin{equation}
\chi (0,0)=\frac{N(0)}{ 1+F_0^a },
\end{equation}
In approaching the quantum critical point from the paramagnetic side $F_0^a$
approaches the value $-1$ from above, while approaching from the
ferromagnetic side $F_0^a$ approaches $-1$ from below. The different sign of 
$1-N(0)V_c\sim 1+F_0^a$ has dramatic effect on the sign of the
spin-fluctuation mediated quasiparticle interaction on the singlet channel
which can be seen in the $t$-matrix\cite{Berk and Schrieffer66} 
\begin{equation}
t(0,0)=V_c+\frac{N^{-1}(0)(F_0^a)^2}{1+F_0^a}\text{.}
\end{equation}
In the paramagnetic phase the second term is positive, while in the
ferromagnetic it is negative leading to $s$-wave pairing.

What is the physics in the neighborhood of the
quantum critical point in weak ferromagnetic metals is still an open
question. Another interesting point is that the BCS theory of
superconductivity can not give a quantum critical point, because as the
critical temperature approaches zero so does the pairing interaction.
Whether a different type of superconductivity exists or a different phase
exists in the neighborhood of the quantum critical point on the paramagnetic
side of the phase diagram is still an open question.
\begin{figure}[h]
\vspace{0.3cm}
\epsfxsize=7.0cm
\hspace*{0.2cm}
\epsfbox{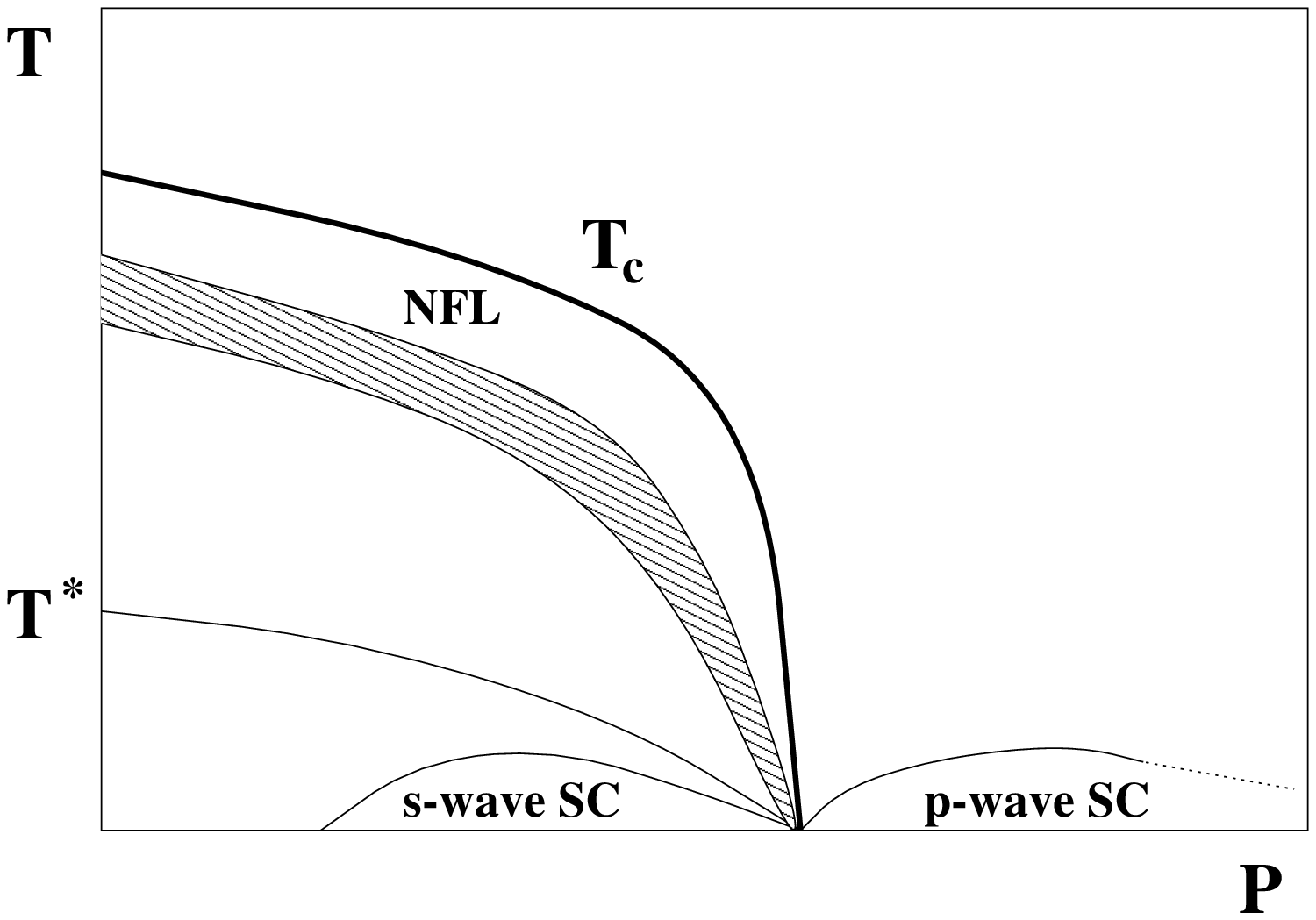}
\caption{The phase diagram of a weak ferromagnetic metal. The non-Fermi liquid
crossover region and the scale set by the temperature $T^*=T_c^2/\epsilon_F$
are explained in the text.}
\label{fig}
\end{figure}
In Fig.(1) we represent a schematic phase diagram of a ``typical'' weak ferromagnetic
metal. Because, the energy scale $T^*=T_c^2/\epsilon_F$ 
below which the superconducting instability occurs vanishes the $s$-wave 
superconducting state must also vanish at the quantum phase transition.
On the paramagnetic side the $p$-wave supercondicting state is expected
as predicted by the DBS theory. At finite temperatures close
to the ferromagnetic phase transition on the ferromagnetic
side the spin fluctuations renormalize the physical quantities leading to
a non-Fermi liquid specific heat $C\sim TlnT$ and in Fig.(1) we have
shown the crossover between the Fermi liquid and the non-Fermi liquid
state. We also expect the $p$-wave superconducting state in the
paramagnetic phase to vanish
at the quantum critical point. Another possibility is that it remains
finite as we go through the phase transition at finite Curie temperature.
However, the superconducting transition temperature must be less
than the Fermi liquid scale set by $T^*$ which vanish
at the quantum phase transition and this implies the vanishing of the
$p$-wave paired state.
  
The $s$-wave superconducting state in the ferromagnetic phase is unusual
and is a generalization of the Larkin-Ovchinnikov-Fulde-Ferrell (LOFF)\ state
\cite{Larkin and Ovchinnikov64,Fulde and Ferrell64} studied in the 60's in
metals with magnetic impurities. The difference between this generalized
LOFF state and the one originally studied is that the magnetic moment in the
former is caused by the quasiparticles which also participate in the
pairing, while in the latter the magnetic field is external to the
quasiparticle system. Therefore the response of the two systems to an
external magnetic field must be quite different. The understanding of how
the spin fluctuations are modified by the superfluid density is an
interesting question which can shed light on the nature of this state.
Another interesting possibility is that this state has an odd-gap
close to half filling induced by the presence of magnon excitations.
The details of this state are beyond the scope of the current paper and
will be investigated in a future publication.

In conclusion, in this paper we described the physics of a weak
ferromagnetic metal from microscopic principles. We have shown that the
vertex corrections in the physical limit are small and that the self-energy
is local. In the limit of small momentum transfer the vertex function
enhances the effective coupling between the quasiparticles in the
neighborhood of the quantum phase transition leading to an $s$-wave
superconducting instability.

We would like to thank J.L. Smith, A. Balatsky, 
P. Coleman, G. Kotliar, A.E. Ruckenstein,
L.B. Ioffe, G.G. Lonzarich, R.B. Laughlin, and W. Mao for the stimulating
discussions on this subject. This work was sponsored by the DOE Grant
DEFG0297ER45636.

\end{document}